\documentclass{aastex61}

\submitjournal{ApJ}

\begin{document}

\title{
High-frequency quasi-periodic light variations from arc-shaped gas clouds falling to a black hole
}

\author[0000-0002-0786-7307]{Kotaro Moriyama}
\affil{Department of Astronomy, Kyoto University, Kitashirakawa, Oiwake-Cho, Sakyo-ku, Kyoto 606-8502, Japan}
\email{moriyama@kusastro.kyoto-u.ac.jp}
\author{Shin Mineshige}
\affiliation{Department of Astronomy, Kyoto University, Kitashirakawa, Oiwake-Cho, Sakyo-ku, Kyoto 606-8502, Japan}

\author{Hiroyuki R. Takahashi}
\affiliation{Center for Computational Astrophysics, National Astronomical Observatory of Japan, Mitaka, Tokyo 181-8588, Japan}



\begin{abstract}
We investigate dynamical and radiative properties of arc-shaped gas clouds falling onto a stellar mass black hole 
based on the three-dimensional general relativistic radiation-magnetohydrodynamics (3D-GRRMHD) simulation data. 
Assuming that the gas clouds radiate due mainly to the free-free emission 
and/or optically thick, inverse Compton scattering, 
we calculate how the emissivity distributions develop with time. 

We find that (1) gas clouds, each of which has a ring-like or arc shape, are intermittently formed, 
and that (2) they slowly fall to the black hole, keeping nearly the Keplerian orbital velocity.
These features support the dynamical properties of the gas clouds assumed in the spin measurement method proposed by \citet{Moriyama1}, 
but the radius of the inner edge of the accretion disk is larger than that of the marginally stable orbit (ISCO).

Next, we examine how each gas cloud is observed by a distant observer
by calculating the photon trajectories in the black hole space-time.
The luminosity of the accretion flow exhibits significant time variations on different timescales, 
reflecting the time evolution of the gas density distributions.
The relatively slow variations on the time durations of $0.08-0.10$ sec is due to the formation and 
fall of gas clouds, while quasi-periodic flux peaks with short time intervals ($0.01$ sec) are
due to the quasi-periodic enhancement of light from the non-axisymmetric arc-shaped clouds 
through the beaming effect. 
This may account for the high-frequency quasi-periodic oscillations (HF QPOs) observed in black hole binaries.
The observational implications and future issues are briefly discussed.
\end{abstract}


\section{Introduction}
General relativity (GR) is the most widely accepted theory of gravitational fields, 
and its validity is tested by the recent observations of the gravitational waves from merging binary black hole systems
(\citealp{Abbott1a}; \citealp{Abbott1b}) and number of observations in the weak field regime (\citealp{Will1}).
The observational determination of the black hole space-time is one of
the most important issues of the general relativity theory, since it can lead to the proof of the event horizon
and the measurement of components of the metric tensor of the space-time.
It is known in general relativity that the black hole space-time is completely determined by the only
two parameters: a black hole mass, $M$, and spin parameter, $a$,
where $a = J/M$, and $J$ is the angular momentum of the black hole 
(we hereafter take the speed of light, $c$, and the gravitational
constant, $G$, to be unity, and neglect the electric charge of black holes, since they seem 
never to be important in the astrophysical context). 
The black hole masses can be relatively easily measured by observing
the motions of stars or gas, since the observed targets may not necessarily be close to the black hole
(\citealp{Shahbaz1}; \citealp{Ghez1}; \citealp{Orosz1}).
However, the measurement of the spin is not easy, 
since its effect is only detected in the vicinity of the black hole, where full consideration 
of general relativistic effects is necessary (see \citealp{Moriyama1}, hereafter Paper I).

In paper I, we proposed a new method for spin measurement based 
on the non-periodic flux variation from an infalling gas blob, ring, or arc-shaped blob.
There we postulated the following features of the gas cloud:
(1) It has a ring or arc shape and is intermittently formed in the innermost region of an accretion disk.
(2) It has nearly the Keplerian orbital velocity and slowly falls to the black hole.
(3) The light variation from the gas cloud is significantly affected by the relativistic effects, 
such as the focusing effect around the photon circular orbit (see paper I, subsection 3.1). 
By means of global magnetohydrodynamics (MHD) simulations of the gas flow in a pseudo Newtonian potential (\citealp{Paczynski1}), 
Machida \& Matsumoto (2003) have proved the feature (1).
They studied the time evolution of a torus and carefully analyzed the simulation results, 
finding that gas clouds with the spiral shape are intermittently formed and fall to the black hole
from the inner edge of the torus. 
It is, however, unclear whether the key features (1)-(3) can be reproduced in the general relativistic regime. 

In order to examine the realistic behavior of the accretion flow near the black hole, 
it is essential to take into account a complex and sensitive interaction between general relativistic, radiation, and magnetic fields.
This requirement is satisfied by the implementation of the three-dimensional general relativistic radiation-magnetohydrodynamics 
(3D-GRRMHD) simulation (\citealp{McKinney1}; \citealp{Mishra1}; \citealp{Takahashi1}; \citealp{Sadowski1}).
We use the latest simulation data of Takahashi et al. (2016), and examine whether the key features assumed in paper I are reproduced.

We analyze the time development of the accretion flow around a stellar mass black hole 
($M=10M_{\odot}$, where $M_{\odot}$ is the solar mass),
and carefully examine how they are observed by a distant observer. 
Here, we consider the accretion flow from the relatively cold disk ($\gtrsim 10^7$K), 
which is truncated near the black hole.
We calculate the light variation from the accretion flow by using the general relativistic ray-tracing method.

Furthermore, the examination of the accretion flow in the vicinity of the black hole enables us to 
explore the origin of the high-frequency quasi-periodic oscillations 
(HF QPOs; the frequency is $\nu_{\rm HF}=100-450$ Hz), 
which have been detected in quite a few black hole low-mass X-ray binaries (\citealp{Remillard4a}).
They are thought as the relativistic phenomenon occurring in the innermost part of the accretion disk,
since $\nu_{\rm HF}^{-1}$ is on the same order of the rotation period at the marginally stable orbit.
The HF QPOs mainly occur in the very high state (the steep power-law state) with sub-Eddington luminosity, 
and their frequencies do not shift freely in response to luminosity changes (\citealp{Remillard1}; \citealp{Remillard4b}). 
In order to explain the observed HF QPOs, a number of models have been proposed (\citealp{Kato1}; \citealp{Rezzolla1}; \citealp{Remillard2}; \citealp{Kato2}; \citealp{KatoY1}), although there is no widely accepted one. 
By using the GR simulation data, we propose an extended physical model 
of a simple hotspot one (\citealp{Schnittman1}; \citealp{Beheshtipour1}) to explain the HF QPOs.

The plan of this paper is as follows: in section 2, we describe our methods 
of calculating the observed flux variation from the simulated accretion flow. 
In section 3, we will show the results and inspect the dynamical and radiative assumptions of 
our gas ring model for spin measurement made in Paper I.
Section 4 is devoted to discussions of the observational implications and future issues.

\section{Model and methods of numerical calculations}
\begin{figure}\begin{center}\includegraphics[width=10cm]{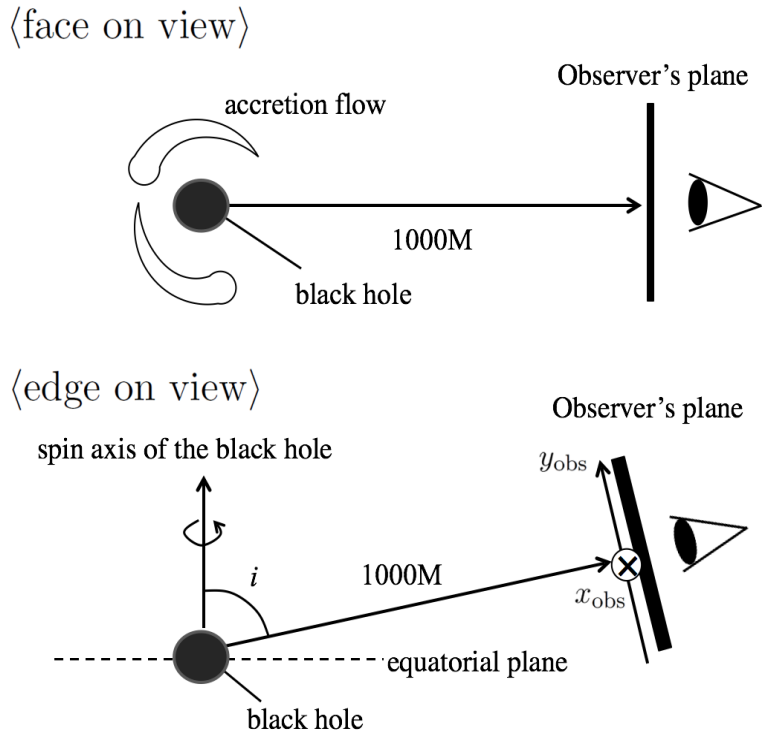} \end{center}
\caption{Schematic picture explaining the motion of the accretion flow and the observer's plane.
Here, $x_{\rm obs}$ and $y_{\rm obs}$ are Cartesian coordinates on the observer's plane, 
the $x_{\rm{obs}}$-axis is parallel to the equatorial plane of the black hole ($\otimes $), 
and the $y_{\rm{obs}}$-axis is perpendicular to the $x_{\rm{obs}}$-axis.
}\label{ray}
\end{figure}
\subsection{Overview of the GRRMHD simulation}
Takahashi et al. (2016) performed the GRRMHD simulation in the polar coordinates
$(t_{\rm KS}, r_{\rm KS}, \theta_{\rm KS}, \phi_{\rm KS})$ in Kerr-Schild space-time with the black hole mass, $M = 10 M_{\odot}$. 
The numbers of the numerical grid points are $(N_{r}, N_{\theta}, N_{\phi})_{\rm KS} = (264, 264, 64)$,
and the computational domain consists of $r_{\rm KS} = [r_{\rm H}, 250M]$, $\theta_{\rm KS} = [0, \pi]$, and $\phi_{\rm KS} = [0,2\pi]$, 
where $r_{\rm H}(=M+\sqrt{M^2 -a^2})$ is the radius of the event horizon. 
They start the simulation from the equilibrium torus given by \citet{Fishbone1}. 
The inner edge of the torus is situated at $r_{\rm KS} = 20M$, while the radius where the pressure reaches its maximum value is $33M$. 
The maximum density of the torus, $\rho_{0}$, is a free parameter and is set to be $10^{-4}\ {\rm g\ cm^{-3}}$. 
They embed weak poloidal magnetic fields inside the torus.
In addition to the torus, they set the dilute, unmagnetized hot atmosphere. 
The density and gas pressure profile of the atmosphere are given by $10^{-4}\rho_{0} (r_{\rm KS}/M )^{-1.5}\ {\rm g\ cm^{-3}}$ 
and $10^{-6} \rho_{0}(r_{\rm KS}/M)^{-2.5}\ {\rm g\ cm^{-1}\ sec^{-2}}$.
Further, they assume a simple $\Gamma$-law for the equation of state: $e = (\Gamma-1)p_{\rm gas}$, 
where $e$ is the inertial energy density, $p_{\rm gas}$ is the gas pressure, and $\Gamma(=5/3)$ is the adiabatic index (see \citealp{Takahashi1}).
We consider the two cases for the black hole spin parameter: $a/M=0$ and $0.9375$. 
In the case of $a/M=0$, we use the simulation data of run A, where the time span is $t_{\rm KS}=0.08-0.23$ sec. 
At $0.08 \leq t_{\rm KS}<0.16$ sec (or $0.16\leq t_{\rm KS}<0.23$ sec), the mass accretion rate is sub-Eddington (super-Eddington); $\dot{M}/L_{\rm Edd}\lesssim 1\ (\sim 2-10)$, where $L_{\rm Edd}$ is the Eddington luminosity (see \citealp{Takahashi1}, section 3.1).
In the case of $a/M=0.9375$, we use the simulation data of run C, where the time span is $t_{\rm KS}=0.08-0.19$ sec.
The mass accretion rate is sub-Eddington ($\dot{M}/L_{\rm Edd}\sim 10^{-1}$) at the whole time (see \citealp{Takahashi1}, section 3.1).
We postulate that the emission region is located within the radial position of the inner edge of the torus; 
that is, $r_{\rm H}/M\leq R_{\rm KS}/M\leq 20$, where $R_{\rm KS}=r_{\rm KS}\sin \theta_{\rm KS}$.
In order to calculate observed radiation from the accretion flow, we transform physical quantities 
in Kerr-Schild space-time into those in Boyer-Lindquist one (see Appendix).
Hereafter, we use the quantities defined by Boyer-Lindquist coordinates.

\subsection{Simple model: calculation of emissivity profiles with no scattering process}
For calculating the emission profiles of the accretion flow, we construct a simple model by making the following  approximations:
\begin{enumerate} 
\item The electron temperature is constant in space, $T_{\rm e}= 10^{7}$ K. 
\item In the comoving frame of a gas element, the emissivity, $j$, is assumed to be proportional to $\rho^2$, 
where $\rho$ is the local gas density.
Here, we postulate that the gas clouds radiate due mainly to the free-free emission
and/or optically thick, inverse Compton scattering (\citealp{Rybicki1}). 
\item We neglect scattering and absorption of radiation in the gas cloud.
\end{enumerate}
In order to obtain the observed light variation from the accretion flow seen by a distant observer, 
we solve the photon trajectories by the ray-tracing method (\citealp{Karas1}; paper I).
The radiative transfer equation is written by
\begin{equation}
\label{eq_jnu}
dI= g^4 j d\ell,
\end{equation}
where $d\ell$ is an infinitesimal spatial interval of a ray in Boyer-Lindquist coordinates, 
and the energy-shift factor, $g$, is expressed as 
\begin{equation}
\label{g}
g = \frac{1}{u^{t} -\Lambda u^{\phi}- V_{r}u^{r}- V_{\theta} u^{\theta}},
\end{equation}
with
\begin{eqnarray}
V_{r} &=&  \frac{1}{\Delta} \sqrt{(r^2+a^2-a\Lambda)^2 -\Delta [(\Lambda -a)^2 +q]},\\
V_{\theta} &=&  \sqrt{q-\cos^2\theta(-a^2+\Lambda^2/\sin^2 \theta)}.
\label{eq_Wetc}
\end{eqnarray}
Here, $\Delta=r^2-2Mr+a^2$, $u^{\mu}$ is the four-velocity of the gas element, 
and $\Lambda$ and $q$ are angular momenta of a ray
with respect to the $\phi$ and $\theta$ directions per unit energy, respectively.
Note that $\Lambda$ and $q$ satisfy
\begin{eqnarray}
\Lambda &=& -x_{\rm obs} \sin i,\\
q&=& y^2_{\rm obs}-a^2 \cos^2 i +\lambda^2 \cot^2 i
\end{eqnarray}
along a ray that reaches a point $(x_{\rm{obs}}, y_{\rm{obs}})$ on the observer's plane (\citealp{Cunningham1}).
Here, $i$ is the inclination angle, and $x_{\rm{obs}}$ and $y_{\rm{obs}}$ are 
Cartesian coordinates on the observer's plane. 
Further, the $x_{\rm{obs}}$-axis is parallel to the equatorial plane of the black hole, and the $y_{\rm{obs}}$-axis is perpendicular to the $x_{\rm{obs}}$-axis (see figure \ref{ray}).

The actual calculation procedures are as follows:
\begin{enumerate}
\item Following Paper I (see subsection 2.4), we calculate the
total intensity of rays, $I = I(x_{\rm obs}, y_{\rm obs}, t)$, that reach each cell at the coordinates of $(x_{\rm obs}, y_{\rm obs})$ at the observer's time of $t$ (see figure \ref{ray}).
We note that equations (7)--(9)  in Paper I must be replaced by equations (\ref{eq_jnu})--(\ref{eq_Wetc}). 

\item We calculate the local flux reaching one cell at $t$, 
$df(x_{\rm obs}, y_{\rm obs}, t) = I(x_{\rm obs}, y_{\rm obs}, t)dS_{\rm obs}/4\pi D^2$, 
where $dS_{\rm obs}$ is the area of the cell, and the distance between 
the center of the observer's plane and black hole, $D$, is set to be $1000M$.
\item In order to obtain the flux, $f(t)$, we integrate $df$ over the entire observer's plane
at each observer's time, $t$,
\begin{equation}
 f(t)=\frac{1}{4\pi D^2}\int I(x_{\rm obs}, y_{\rm obs}, t)dS_{\rm obs}.
\end{equation}
\item We define the normalized flux; $F(t) \equiv f(t)/f(t_{\rm max})$, where $t_{\rm max}$ denotes the time when $f(t)$ reaches its maximum.
\end{enumerate}

In order to analyze the time variation of the gas accretion, 
we need to examine the emissivity distribution on the equatorial plane.
We define the vertical average of the emissivity:
\begin{equation}
\label{var_j}
  \overline{j}(x,y,t)=\frac{1}{4M}\int^{2M}_{-2M}j(t,x,y,z) dz,
\end{equation}
where $x,y,$ and $z$ are Cartesian coordinates transformed from Boyer-Lindquist ones.
Further, we neglect the radiation from the region of $|z/M|>2$, since the emissivity in the region is $10^{-4}$ times 
smaller than the maximum emissivity at each time.

\subsection{Free-free model: contribution of temperature distribution}\label{free}
In the case of the simple model, we assume a constant and uniform electron temperature, $T_e = 10^{7}$ K.
In order to investigate the impacts of this assumption on results, 
we construct another model by considering the temporal and spatial variations of the temperature (free-free model).
We assume a one-temperature plasma in which the electron temperature is equal to the ion temperature obtained by the GRRMHD simulation, 
and consider the free-free emission, i.e., the emissivity is proportional to $\rho^2 T_{\rm e}^{1/2}$.
We note that the other approximations and numerical procedures are the same with those of the simple model.

\subsection{Scattering model: consideration of inverse Compton scattering}
\begin{figure}\begin{center}\includegraphics[width=18cm]{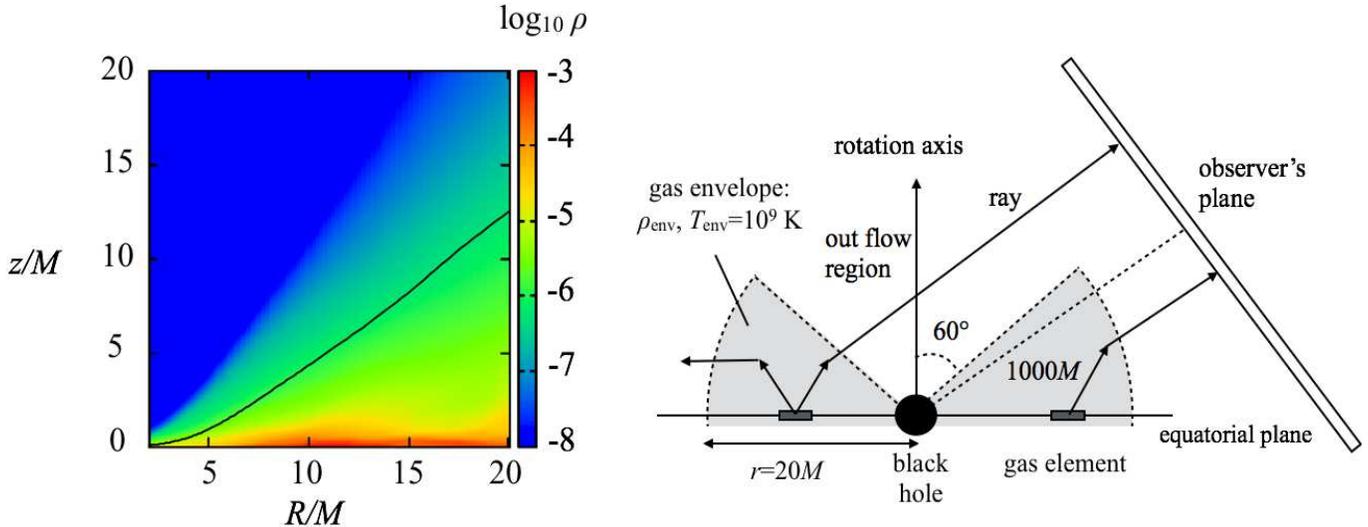} \end{center}
\caption{
Consideration of inverse Compton scattering.
The left panel shows the spatial distribution of the gas density averaged over $0\leq \phi <2\pi$ and $0.12\leq t \leq 0.23$ sec.
The high-density gas cloud locates around the equatorial plane ($|z|/M\lesssim 2$), 
and is surrounded by the low-density gas envelope ($|z|/M\gtrsim 2$).
The solid curve plots the photosphere, where $\tau_{\rm es}=1$ (equation \ref{tau_es}).
In the region $i\lesssim 60^\circ$, $\tau_{\rm es}$ is smaller than $1$, 
since there is the high temperature outflow and its gas density is much smaller than that of the gas envelope.
The right panel depicts the schematic picture of the scattering model.
The radiation is affected by inverse Compton scattering, and then reaches the observer's plane.
We postulate that photons do not undergo the electron scattering in $i< 60^\circ$, since $\tau_{\rm es}<1$.
Here, the radius of the envelope is set to be $20M$, and $\rho_{\rm env}$  
and $T_{\rm env} (=10^9\ {\rm K})$ are the gas density and electron temperature of the envelope, respectively.
}\label{scat_ab}
\end{figure}

We finally construct a scattering model by considering the light variation under inverse Compton scattering and free-free absorption in the case of the non-rotating black hole. 
The electron-scattering optical depth, $\tau_{\rm es}$, and effective optical depth, $\tau_{\rm eff}$, are expressed as
\begin{eqnarray}
\label{tau_es}
\tau_{\rm es} &=& \int^{z_{\rm max}}_{z} \rho \kappa_{\rm es} dz,\\
\tau_{\rm eff} &=& \int^{z_{\rm max}}_{z} \rho\sqrt{(\kappa_{\rm es}+\kappa_{\rm ff})\kappa_{\rm ff}}dz,
\end{eqnarray}
where $\kappa_{\rm es}=0.4 \ {\rm cm^{2}g^{-1}}$, $\kappa_{\rm ff}=6.4\times 10^{22}\rho T_{\rm e}^{-3.5} \ {\rm cm^{2}g^{-1}}$, 
and $z_{\rm max}(=20M)$ is set to be the maximum height of the radiation region.
In order to construct the scattering model, we show in the left panel of figure \ref{scat_ab} 
the spatial distribution of the gas density averaged over $0\leq \phi <2\pi$ and $0.12\leq t \leq 0.23$ sec.
There is the high-density gas cloud ($\rho= 10^{-3}-10^{-5}\ {\rm g\ cm^{-3}}$) in the region of $|z|\lesssim 2M$, 
while the low density gas envelope ($\rho\lesssim 10^{-5}\ {\rm g\ cm^{-3}}$) exists in the region of $|z|\gtrsim 2M$.
The solid curve in the left panel of figure \ref{scat_ab} plots the photosphere, where $\tau_{\rm es}=1$, and roughly corresponds to $i= 60^\circ$. 
In the region $i\lesssim 60^\circ$, $\tau_{\rm es}$ is smaller than $1$, 
since there is the high temperature outflow and its gas density is much smaller than that of the gas envelope.
Thus inverse Compton scattering mainly occurs at $i\gtrsim 60^\circ$.
From these properties of the accretion gas flow, we construct the following model of gas clouds and envelope 
(see the right panel of figure \ref{scat_ab}): 
\begin{enumerate}
\item The electron temperature of the gas cloud region ($|z|/M\leq 2$) is constant, $T_{\rm cloud}=10^7$ K.
\item Gas clouds distribute to the equatorial plane and the initial intensity is given by $I_{0}\propto \overline{j}$.
\item The soft photon is isotropically emitted from the gas cloud region with the energy, $k_{\rm B}T_{\rm cloud}$, 
where $k_{\rm B}(=1.38\times 10^{-16}\ {\rm erg\ K^{-1}})$ is the Boltzmann's constant.
\item In the region of $|z|> 2M$ and $i \geq 60^\circ$, there is the static gas envelope 
with the gas density, $\rho_{\rm env}(<10^{-5}{\rm g\ cm^{-3}})$, and electron temperature, $T_{\rm env}=10^9\ {\rm K}$ (\citealp{Sadowski2a}).
\item  In the region $i<60^\circ$, the radiation does not undergo the electron scattering, since $\tau_{\rm es}<1$.
\end{enumerate}
We note that the free-free absorption is neglected in the envelope region, since the free-free absorption 
$\tau_{\rm eff}\lesssim 10^{-7}$ is much smaller than 1, 
where we set $z=-20M$ as the bottom height of the radiation region.
We approximately calculate the observed intensity, $I$, affected by inverse Compton scattering process (\citealp{Rybicki1}):
\begin{eqnarray}\label{scatter_eq}
I&=& g^4 I_0e^y,
\end{eqnarray}
where $y=4kT_{\rm env}{\rm max}(\tau_{\rm scat},\tau_{\rm scat}^2)/(m_{\rm e}c^2)$ is the Compton y parameter, 
$m_{\rm e}$ is the electron mass, $\tau_{\rm scat}$ is the electron-scattering optical depth along the ray trajectory
and $g$ is the energy-shift factor of the gas cloud (equation \ref{g}).

The procedure of the numerical calculation is as follows:
\begin{enumerate} 
\item In the comoving frame of the gas cloud located at $(x,y)$ at the time $t_{\rm i}$ ($0.12\leq t_{\rm i} \leq 0.23$ sec), we calculate the trajectories of the rays 
by using the symplectic method. 
%
\item We calculate the electron-scattering optical depth, $\tau'_{\rm scat}=\int \rho \kappa_{\rm es}d\ell$, along a ray trajectory, 
where $d\ell$ is the infinitesimal spatial interval of the ray in Boyer-Lindquist coordinates.
At the position where $\tau'_{\rm scat}$ reaches $1$, we reselect the direction of the ray by using a uniform random number, 
assuming isotropic emission in the comoving frame of the gas element.
Then we take $\tau'_{\rm scat}=0$, and calculate $\tau'_{\rm scat}$ along the new ray trajectory.
\item If the ray reaches the observer's plane, we estimate the traveling time, 
$t_{\rm f}$, and other quantities, 
${\rm max}(\tau_{\rm scat},\tau_{\rm scat}^2)$, $g$ and $I_{0}$, where ${\rm max}(\tau_{\rm scat},\tau_{\rm scat}^2)$ is 
given by the number of the scattering of the ray. 
We substitute them to the equation (\ref{scatter_eq}), and calculate the observed intensity emitted from the gas element located at ($x,y$), $dI(x,y,t)$, where $t=t_{\rm f}+t_{\rm i}-D$, 
and $D(=1000M)$ is the distance between the center of the observer's plane and black hole.
\item We perform the upper procedure to the all elements of gas clouds at whole time, 
and then obtain the total intensity, $I_{\rm tot} (t)$, by superposing each intensity profile, $dI(x,y,t)$.
By replacing  the term of "$I$" in equation (7) with "$I_{\rm tot} (t)$", we obtain the observed energy flux, $f(t)$.
%
\item Further, we calculate the Compton y parameter, and then we estimate the y parameter averaged over the all rays reaching the observer's plane, $y_{\rm av}$, 
where we use $dI(x,y,t)$ as the weighted function. 
\end{enumerate}

\section{Results}
\subsection{Results of simple model: overall behavior of accretion flow ($a/M=0$)}

\begin{figure}\begin{center}\includegraphics[width=17cm]{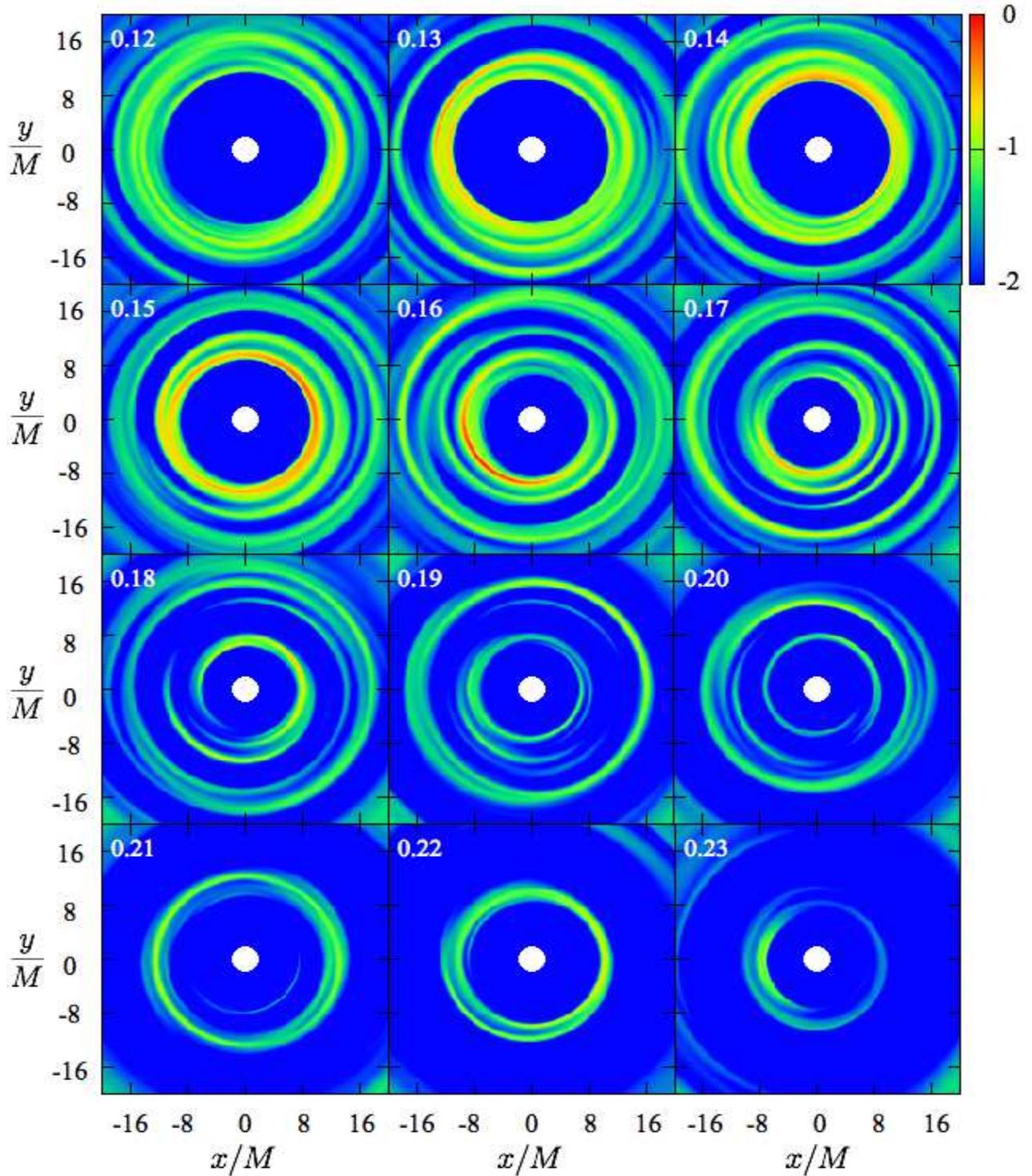} \end{center}
\caption{Time development of the spatial distribution of the vertically averaged emissivity, 
$\overline{j}(x,y,t)$, near the non-rotating black hole.
The color represents the logarithm of $\overline{j}(x,y,t)$ 
normalized by the absolute maximum one, which is found at $(x,y)=(-5.9M,-7.1M)$ at the time $t=0.16$ sec.
The origin is set to be at the center of the black hole, the white circle indicates the event horizon, and the gas cloud is rotating in a counterclockwise direction.
The observer's time, $t$ (sec), is shown in the upper-left corner of each panel.
}\label{jnu}
\end{figure}

\begin{figure}\begin{center}\includegraphics[width=15cm]{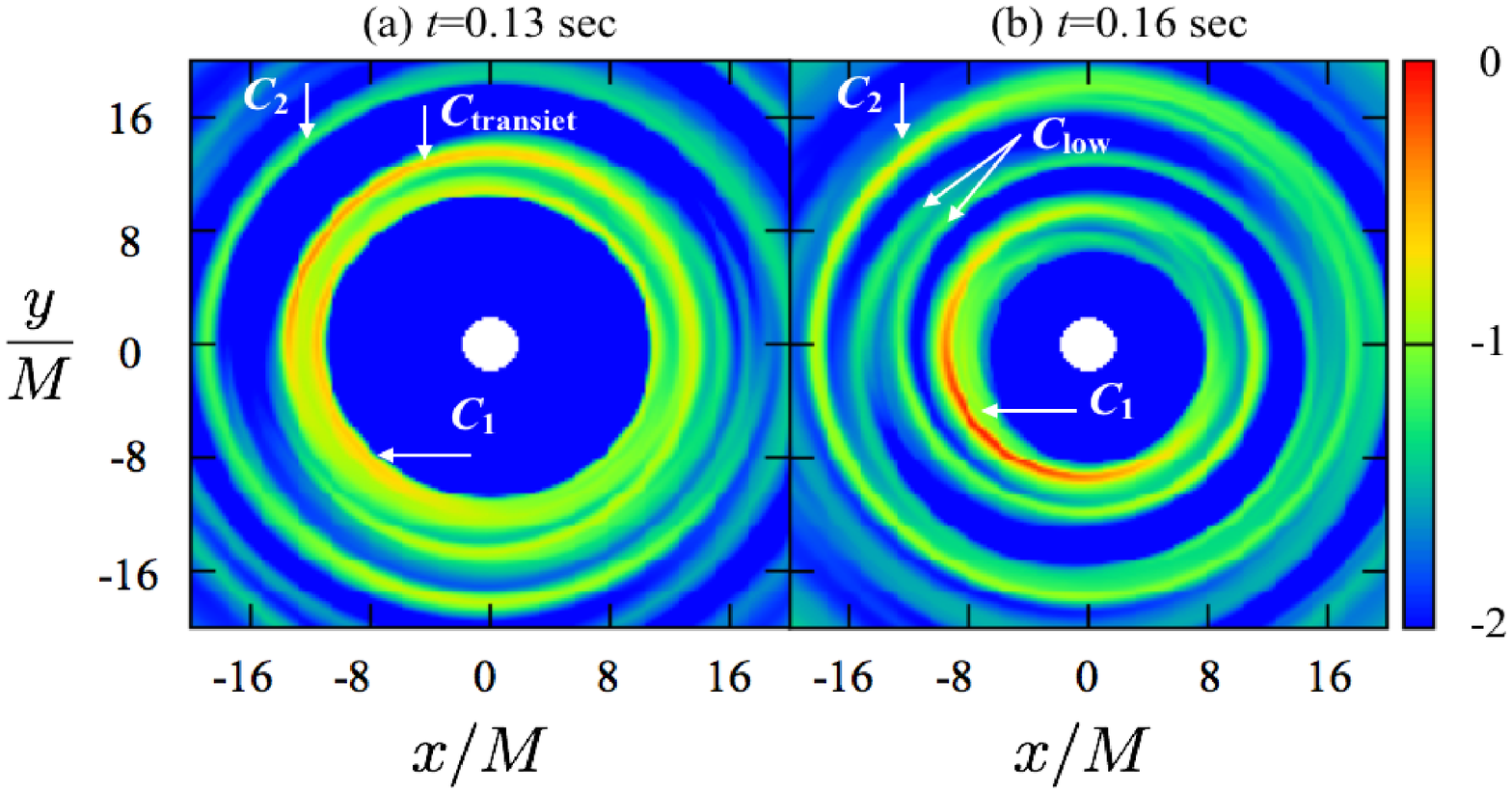} \end{center}
\caption{
Characteristic arc-shaped gas clouds.
The definition of the color is the same as that of figure \ref{jnu}.
Three arc-shaped gas clouds ($C_1$, $C_{\rm transient}$, and $C_{2}$)
are formed at $t=0.13$ sec.
At $t=0.16$ sec, $C_{\rm transient}$ has been already captured by $C_1$, and low-emissivity gas clouds
$C_{\rm low}$ are formed.
}\label{Cs}
\end{figure}

In order to understand the time development of the accretion flow around the non-rotating black hole, 
we show in figure \ref{jnu} the time evolution of the spatial distribution of 
the vertically averaged emissivity during $t=0.12-0.23$ sec, 
where $t$ is the observer's time, and is shown in the upper-left corner of each panel.
The color represents the logarithm of $\overline{j}(x,y,t)$ 
normalized by the absolute maximum one, ${\rm max}\left[\overline{j}(x,y,t)\right]$, which is found at $(x,y)=(-5.9M,-7.1M)$ at the time $t=0.16$ sec.

We confirm that arc-shaped gas clouds are intermittently formed, and fall onto the black hole:
\begin{enumerate}
\item At $t=0.12$ sec, an arc-shaped gas cloud is formed at $R/M=14$, and gradually falls to the black hole. 
At $t=0.13$ sec, the gas cloud grows and separates into two arc-shaped gas clouds ($C_{1}$ and $C_{\rm transient}$), 
and a low-emissivity gas cloud, $C_{2}$, is formed outside $C_{\rm transient}$ (see figure \ref{Cs}a).
\item
During $0.13-0.16$ sec, $C_{\rm transient}$ is captured by $C_{1}$.
In addition, low-emissivity gas clouds, $C_{\rm low}$, are formed around $C_1$ at $t=0.16$ sec (see figure \ref{Cs}b) 
and are captured by $C_{1}$ and $C_{2}$ during $0.16-0.20$ sec. 
\item At $t=0.19$ sec, the emissivity of $C_{1}$ decays, while that of $C_{2}$ increases.
During $t=0.20-0.23$ sec, $C_2$ falls to the black hole. 
\end{enumerate}
The density fluctuation of $C_1\ (C_2)$ occurs during $0.12-0.21\ (0.13-0.23)$,
and so the lifetime is $0.09\ (0.10)$ sec.
During $0.13-0.18\ (0.20-0.23)$ sec, the major radiation is emitted from $C_1\ (C_2)$ which slowly falls to the black hole.

\begin{figure}\begin{center}\includegraphics[width=12cm]{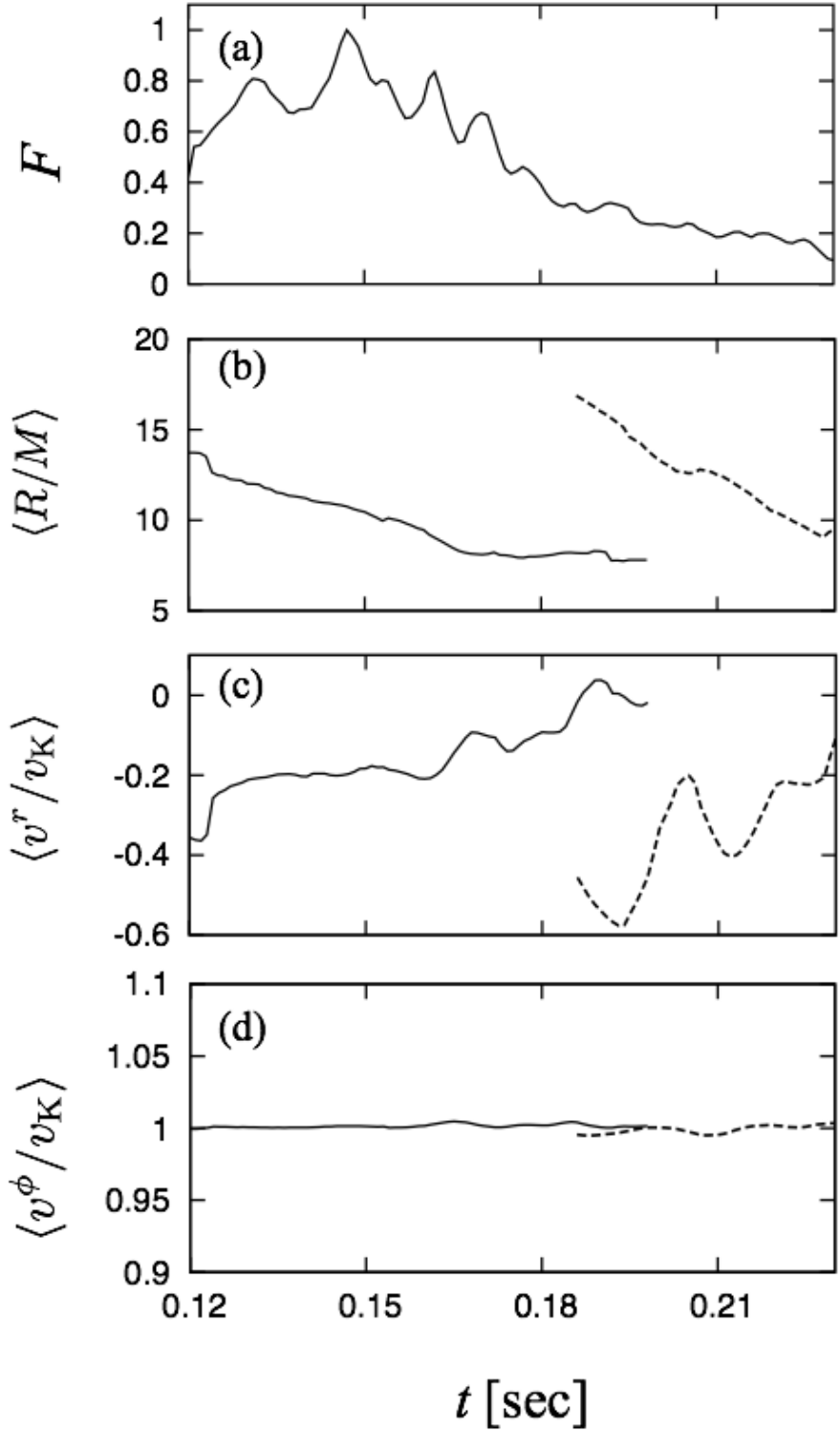} \end{center}
\caption{Radiative and dynamical properties of the accretion flow around the non-rotating black hole: 
(a) the time variation of the normalized flux of the accretion gas cloud for the case of $i=60^{\circ}$, 
(b)--(d) time variations of the volume averaged radial position, $R$, and velocities, $v^{\mu}=dx^{\mu}/dt$.
Here, we use the emissivity as the weighted function, and $v_{\rm K}$ denotes the Keplerian orbital velocity.
The solid (dashed) curve indicates the value of the arc-shaped gas cloud, $C_1\ (C_2)$.
}\label{light_u4u}
\end{figure}

We show in figure \ref{light_u4u}a the light curve of the accretion flow for the case with $(a/M, i) = (0,60^\circ)$.
The flux variation can be divided into two stages:
\begin{enumerate}
\item After the formation of the first gas cloud at $t=0.12$ sec, the flux first increases until the flux maximum at $t =0.147$ sec with the growth of $C_1$. 
\item The flux then decays during $0.15-0.20$ sec, since $C_1$ falls to the black hole and its emissivity decreases. 
The flux gradually decreases during $0.20\leq t \leq 0.23$ sec, since $C_2$ falls to the black hole. 
\end{enumerate}

In order to understand the dynamical properties of the arc-shaped gas clouds, we calculate how the radial coordinates and velocity components, $v^{\mu}(=dx^{\mu}/dt)$, of the gas clouds vary with time.
Here, we focus on the arc-shaped gas clouds, $C_{1}$ and $C_{2}$, 
and do not examine further properties of $C_{\rm transient}$ and $C_{\rm low}$ in this study, 
since their formations are transient. 
We define the following average of the physical quantity, $G(=R,v^{\mu})$: 
\begin{equation}
\langle G\rangle =\int G j d^3 x\bigg/\int jd^3 x,
\end{equation}
where we perform the volume integration over $R_{\rm s}-2M\leq R\leq R_{\rm s}+2M$, $-2M\leq z\leq 2M$, and $0\leq \phi\leq 2\pi$.
Further, $R_{\rm s}(t)$ is defined as the radial position where $\tilde{\rho}(R,t)[=\int\int \rho (R,\phi,z)Rd\phi dz /\int\int Rd\phi dz]$ reaches its maximum at time, $t$; that is, it represents the radial position of the arc-shaped gas cloud.

We show in figures \ref{light_u4u}b--\ref{light_u4u}d the time developments of $R$ and $v^{\mu}$, 
where $v_{\rm K} [=M^{1/2}/(r^{3/2}+aM^{1/2})]$ is the Keplerian orbital velocity,  
and the solid (or the dashed) curve in each panel indicates the value of $C_{1}\ (C_{2})$.

The time variations of the radial positions of the gas clouds, $\langle R\rangle$, indicate that the gas clouds slowly fall to the black hole (figure \ref{light_u4u}b).
During $0.12- 0.19$ sec, $\langle R\rangle$ of $C_1$ gradually decreases, since $C_{1}$ slowly falls to the black hole.
During $0.19- 0.23$ sec, $C_{2}$ grows and then falls to the black hole, 
and so $\langle R\rangle$ of $C_{2}$ decreases.
From figure \ref{light_u4u}c, we also find that the radial velocities of gas clouds, $\langle v^{r}\rangle$, 
are negative and small ($-0.01<\langle v^{r}\rangle< 0.001$); 
that is, each cloud gradually falls to the black hole (see figure \ref{light_u4u}c). 
Remarkably, figure \ref{light_u4u}d proves that the azimuthal angular velocities are very close to 
$v_{\rm K}$ ($0.994< \langle v^{\phi}/v_{\rm K}\rangle < 1.005$), 
while the polar angular velocities are smaller than one percent of $v_{\rm K}$ 
($-0.006< \langle v^{\theta}/v_{\rm K}\rangle < 0.010$).
Therefore each gas cloud has nearly the Keplerian orbital velocity, 
but slowly falls to the black hole with a small radial velocity.
These features justify the basic assumptions made in Paper I.

\subsection{Short-term light variation of arc-shaped gas cloud $(a/M=0)$}
\begin{figure}\begin{center}\includegraphics[width=19cm]{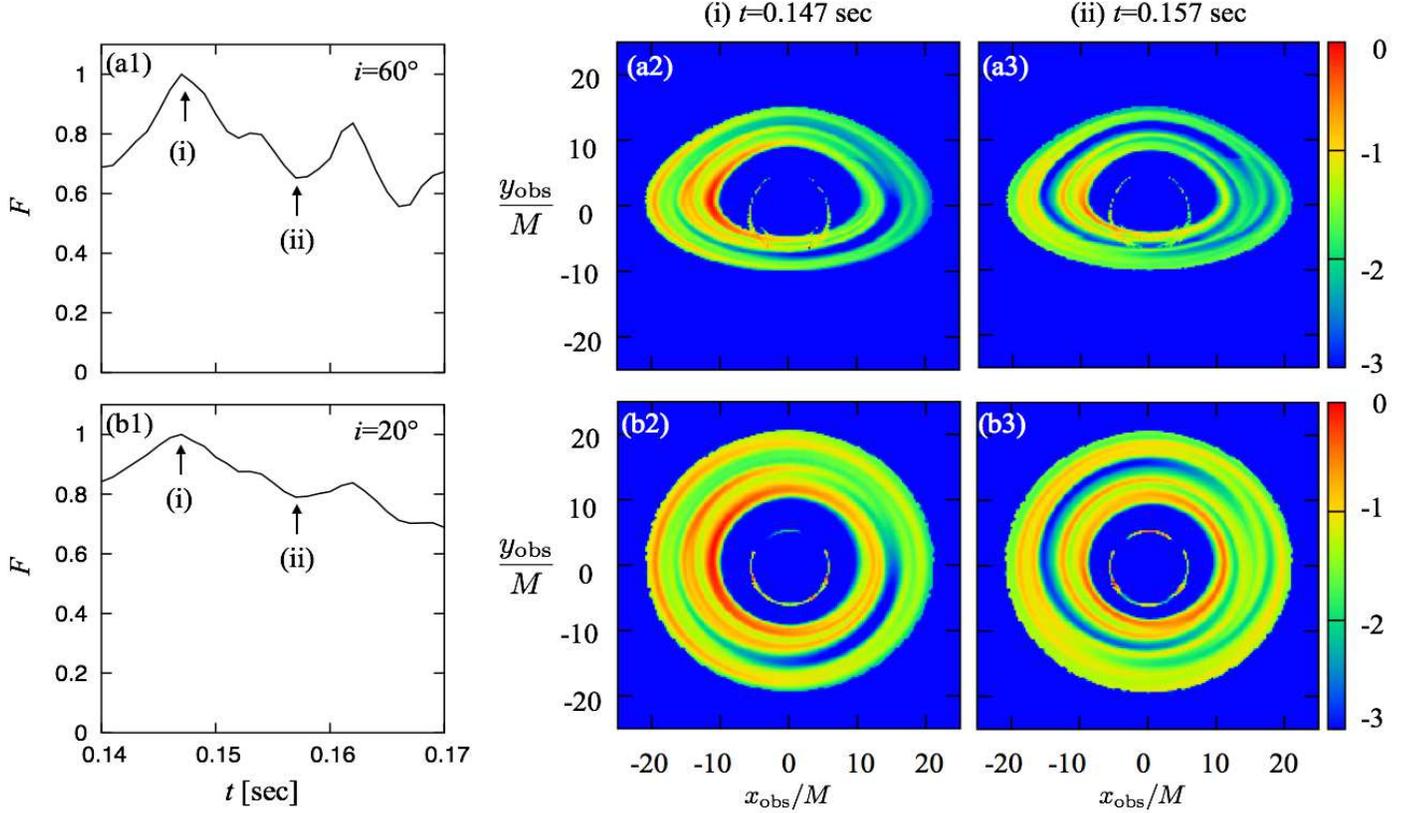} \end{center}
\caption{Origin of the short-term light variation. 
Panel a1 shows the detailed light curve around the typical peak ($t=0.147$ sec) for the case of $i=60^\circ$. 
Panels a2 and a3 are the snapshots at the time of the peak (i) and valley (ii), respectively.
Here, the abscissa is $x_{\rm obs}/M$, while the ordinate is $y_{\rm obs}/M$. 
The color represents the logarithm of the relative brightness of the gas cloud image: 
$\log_{10}\left[df(x_{\rm obs}, y_{\rm obs}, t)/{\rm max}(df(x_{\rm obs}, y_{\rm obs}, t))\right]$, 
where the maximum local flux, ${\rm max}(df(x_{\rm obs}, y_{\rm obs}, t))$, 
is located at $(x_{\rm obs}, y_{\rm obs})=(-11.3M,0.3M)$ at the time $0.147$ sec.
Panels b1-b3 are the same as panels a1-a3, but for the low inclination angle case, $i=20^{\circ}$, 
where the maximum local flux, ${\rm max}(df(x_{\rm obs}, y_{\rm obs}, t))$, 
is located at $(x_{\rm obs}, y_{\rm obs})=(-11.3M,0.3M)$ at the time $t=0.147$ sec.
}\label{light}
\end{figure}

In addition to the slow variations, the light curve has many peaks with short time intervals (see figure \ref{light_u4u}a).
In order to understand the physical origin producing such peaks, 
we focus on the typical flux peak (i) and valley (ii) in panel a1 of figure \ref{light}.
We show the snapshots of the emissivity images at the time of the peak (i)
and valley (ii) in figures \ref{light}a2 and \ref{light}a3, respectively.
The color represents the logarithm of the relative brightness of the gas cloud image: 
$\log_{10}\left[df(x_{\rm obs}, y_{\rm obs}, t)/{\rm max}(df(x_{\rm obs}, y_{\rm obs}, t))\right]$, 
where the maximum local flux, ${\rm max}(df(x_{\rm obs}, y_{\rm obs}, t))$, 
is located at $(x_{\rm obs}, y_{\rm obs})=(-11.3M,0.3M)$ at the time $0.147$ sec.
In panel a2 (or a3), the relative brightness shows its local maximum (minimum) due to the enhancement of the light 
from the non-axisymmetric arc-shaped clouds through the beaming effect, 
since the flux from the gas cloud that is moving towards (away from) the observer is strongly Doppler boosted
(de-boosted) near the black hole (see \citealp{Moriyama2} and references therein).
We confirm the same features for other peaks and valleys, and conclude that 
the physical origin producing the peaks is the beaming effect of the non-axisymmetric arc-shaped gas cloud.
Panels a2 and a3 show that there are three bright regions due to the beaming effect from $C_1, C_2,$ and $C_{\rm transient}$.
We note that the dominant radiation is emitted from the innermost region corresponding to $C_1$.
During $0.13-0.19\ (0.20-0.23)$, the major radiation is emitted from $C_1\ (C_2)$ (see subsection 3.1), 
and so the averaged peak interval of $C_1\ (C_2)$, is $1.0\times 10^{-2}\ (6.6 \times 10^{-3})$ sec.

We also examine the low-$i$ case (say, $i=20^{\circ}$), and show the results in panels b2 and b3.
We find similar tendencies but that the variation amplitudes are smaller than that in the case of $i=60^{\circ}$. 
This can be understood, since it shows that the peak amplitudes are mainly determined by the beaming effect (see panels a1 and b1).
Panel b2 (b3) also shows that the radiation from arc-shaped gas clouds is Doppler boosted (de-boosted).
\subsection{Spiral gas clouds in high-spin case $(a/M=0.9375)$}

\begin{figure}\begin{center}\includegraphics[width=17cm]{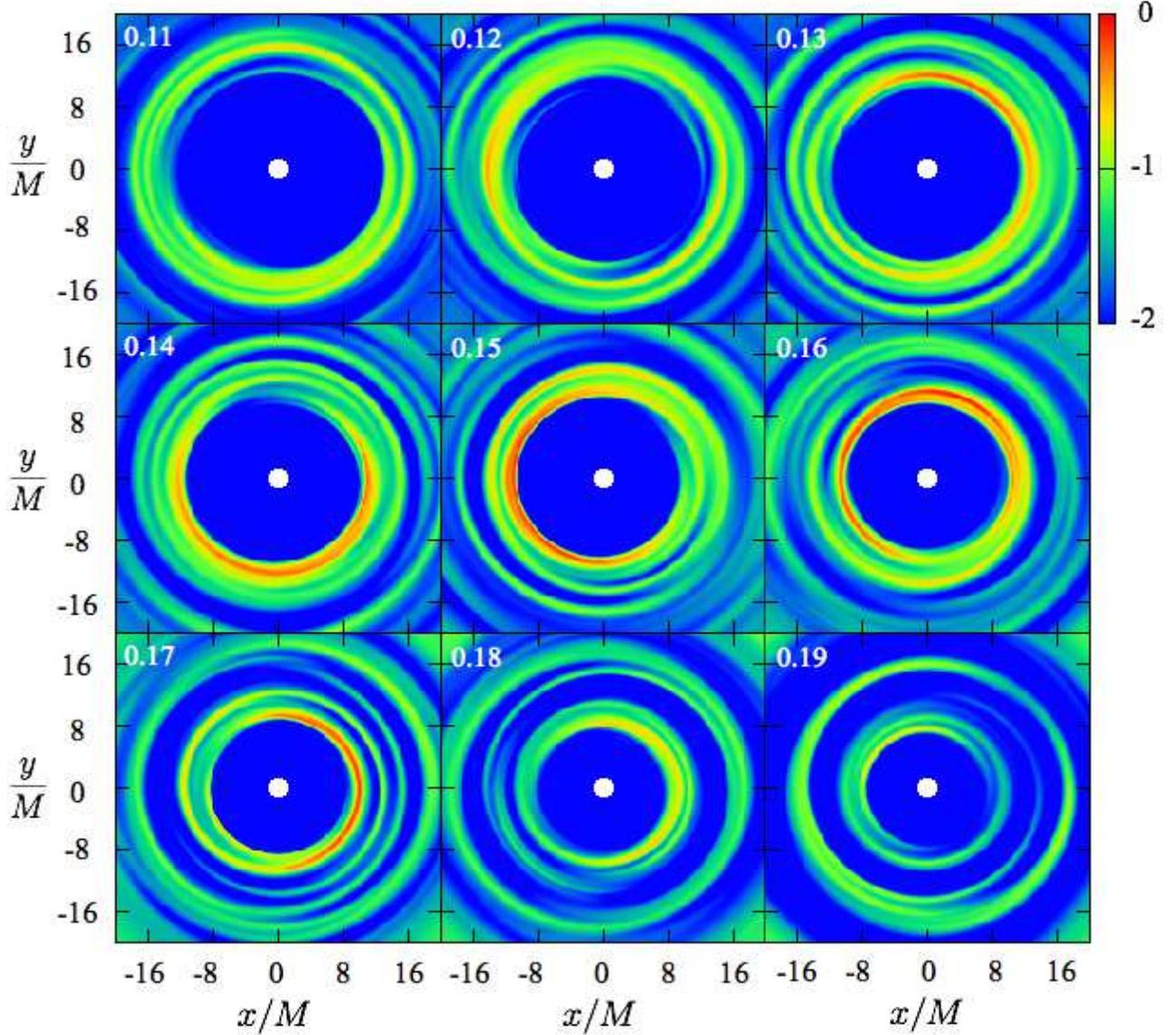} \end{center}
\caption{The same as figure \ref{jnu}, but for the high-spin case, $a/M=0.9375$.
Here, $\overline{j}(x,y,t)$ reaches its absolute maximum at $(x,y,z,t)=(5.5M,9.9M,0.16{\ \rm sec})$.
}\label{jnu_high}
\end{figure}
\begin{figure}\begin{center}\includegraphics[width=17cm]{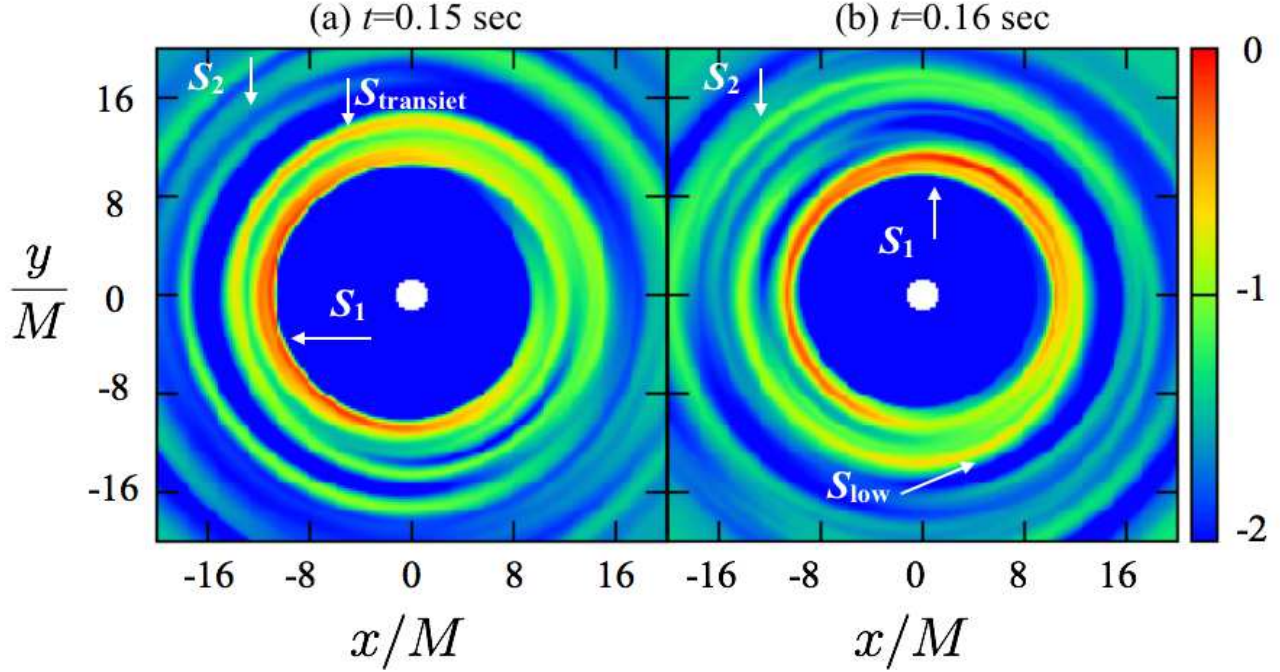} \end{center}
\caption{
Characteristic arc-shaped gas clouds for the high-spin case, $a/M=0.9375$.
The definition of the color is the same as that of figure \ref{jnu_high}.
Three arc-shaped gas clouds ($S_1$, $S_{\rm transient}$, and $S_{2}$)
are formed at $t=0.15$ sec.
At $t=0.16$ sec, $S_{\rm transient}$ has been already captured by $S_1$, and low-emissivity gas clouds
$S_{\rm low}$ are formed.
}\label{Ss}
\end{figure}

We show in figure \ref{jnu_high} the time evolution of the spatial distribution 
of the vertically averaged emissivity in the high-spin case ($a/M=0.9375$). 
We also confirm that arc-shaped gas clouds are intermittently formed and fall onto the black hole:
\begin{enumerate}
\item At $t=0.11$ sec, first arc-shaped gas cloud, $S_1$, is formed at $R/M=14$, 
and gradually falls to the black hole. 
At $t=0.15$ sec, other gas clouds ($S_{\rm transient}$ and $S_{\rm 2}$) are formed outside $S_1$ (figure \ref{Ss}a), 
and then $S_{\rm transient}$ is captured by $S_1$ at $0.16$ sec (figure \ref{Ss}b).
\item During $0.16-0.19$ sec, there are three arc-shaped gas clouds ($S_1, S_{\rm low},$ and $S_{2}$, see figure \ref{Ss}b), 
and $S_1$ is the brightest gas cloud.
\end{enumerate}
At the whole time, the major radiation is emitted from $S_1$ and its emissivity decreases after $0.16$ sec.
The density fluctuation of $S_1$ occurs during $0.11-0.19$ sec, and so its lifetime is $0.08$ sec.

\begin{figure}\begin{center}\includegraphics[width=11cm]{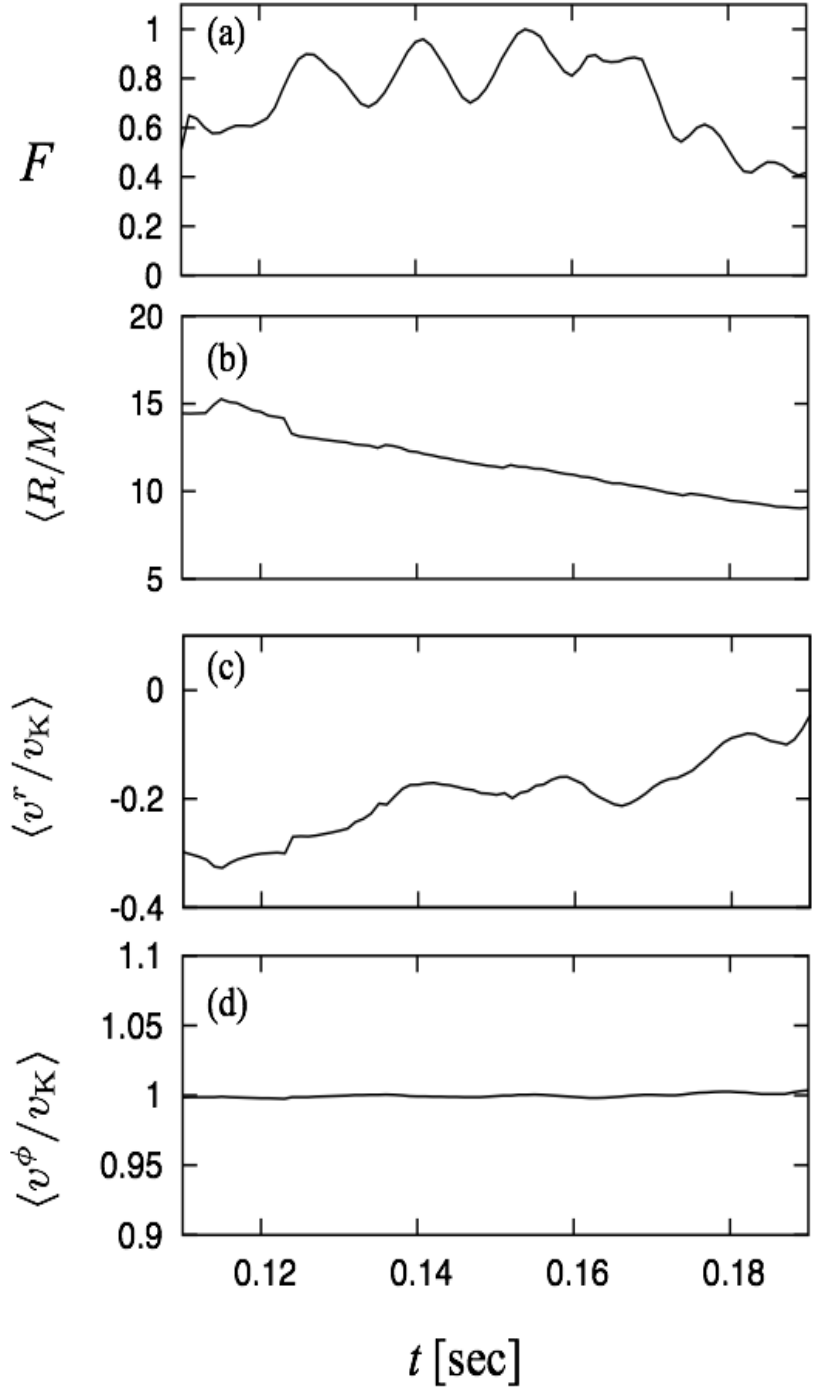} \end{center}
\caption{The same as figure \ref{light_u4u}, but for the high-spin case, $a/M=0.9375$.
}\label{light_u4u_high}
\end{figure}

Next, we show in figure \ref{light_u4u_high}a the light curve of the accretion flow in the case of $i=60^\circ$.
The flux variation can be divided into two stages:
\begin{enumerate}
\item After the formation of $S_1$ at $t=0.11$ sec, the flux first increases until the flux maximum at $t =0.154$ sec with the growth of $S_1$. 
\item The flux then decays during $0.154-0.19$ sec, since $S_1$ falls to the black hole and its emissivity decreases. 
\end{enumerate}

The time developments of the radial coordinate and $v^{\mu}$ of the arc-shaped gas cloud, $S_{1}$, have similar behaviors of those for the non-rotating black hole case (figure \ref{light_u4u_high}b-d). 
Here, we analyze the dynamical property of $S_{1}$, 
and do not examine further properties of $S_2$, $S_{\rm transient}$, and $S_{\rm low}$, 
since their formations are transient, and the major radiation is emitted from $S_1$. 

During $t=0.11-0.19$ sec, $\langle R\rangle$ of $S_1$ gradually decreases, and the radial velocity, 
$\langle v^{r}\rangle$, is negative and small ($-0.005< \langle v^{r}\rangle < -0.001$), 
since it gradually falls to the black hole (see figures \ref{light_u4u_high}b and \ref{light_u4u_high}c). 
We show in figure \ref{light_u4u_high}d that the azimuthal angular velocity is very close to the Keplerian orbital velocity including the spin dependency, ($0.997< \langle v^{\phi}/v_{\rm K}\rangle < 1.004$), 
while the polar angular velocity is smaller than one percent of the Keplerian one ($-0.005< \langle v^{\theta}/v_{\rm K}\rangle < 0.002$).
Therefore, $S_1$ has nearly the Keplerian orbital velocity, but slowly falls to the black hole with a small radial velocity.

\begin{figure}\begin{center}\includegraphics[width=19cm]{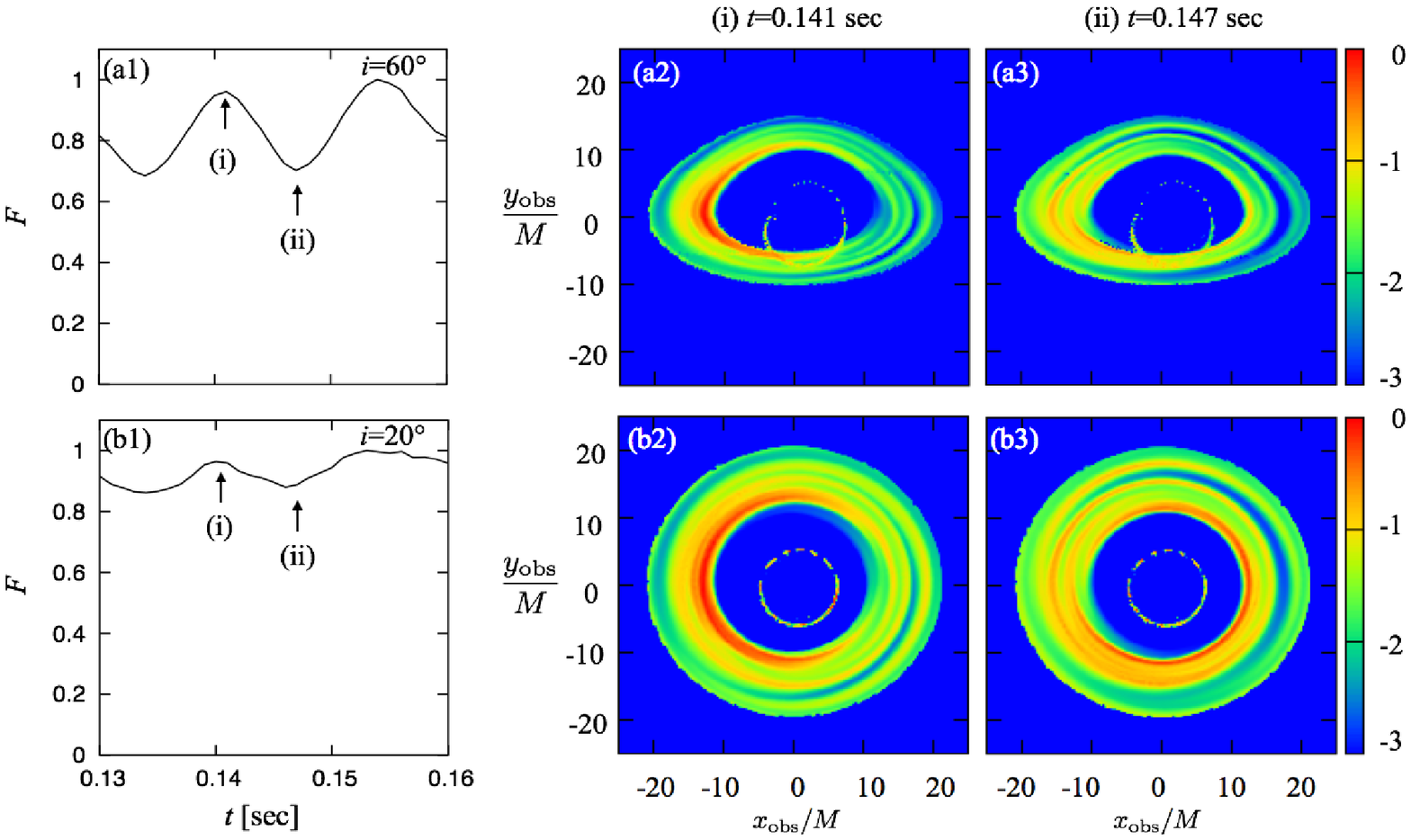} \end{center}
\caption{The same as figure \ref{light}, but for the high-spin case, $a/M=0.9375$. 
Here, the maximum local flux is ${\rm max}(df(x_{\rm obs}, y_{\rm obs}, t))= df(-13.0M,0.3M,0.141)\ [df(-13.0M,0.5M,0.141)]$ in the case of $i=60^{\circ}\ (20^{\circ})$.
}\label{light_high}
\end{figure}

As is the case of the non-rotating black hole, the light curve has many peaks with short time intervals, 
and its origin is the beaming effect of $S_1$.
The averaged peak interval is $9.7\times 10^{-3}$ sec.
In order to confirm this fact, we show in figure \ref{light_high} the light curve around the typical peak (i) and 
the snapshots at the time of the peak (i) and valley (ii), where $i=60^{\circ}$ (panels a1--a3) 
or $20^{\circ}$ (panels b1--b3). 
By carrying out the same discussion for the non-rotating case (see subsection 3.2), 
we conclude that the physical origin producing the peaks is the beaming effect of the non-axisymmetric gas cloud, 
and the peak amplitudes are mainly determined by the beaming effect.
The dominant radiation at the time of peak is emitted from $S_1$, 
and the other radiation from $S_{\rm transient}$ and $S_{\rm low}$ is smaller than that from $S_1$ (panel a2 and b2).

\subsection{Results of free-free model: temperature dependence of gas clouds}
\begin{figure}\begin{center}\includegraphics[width=14cm]{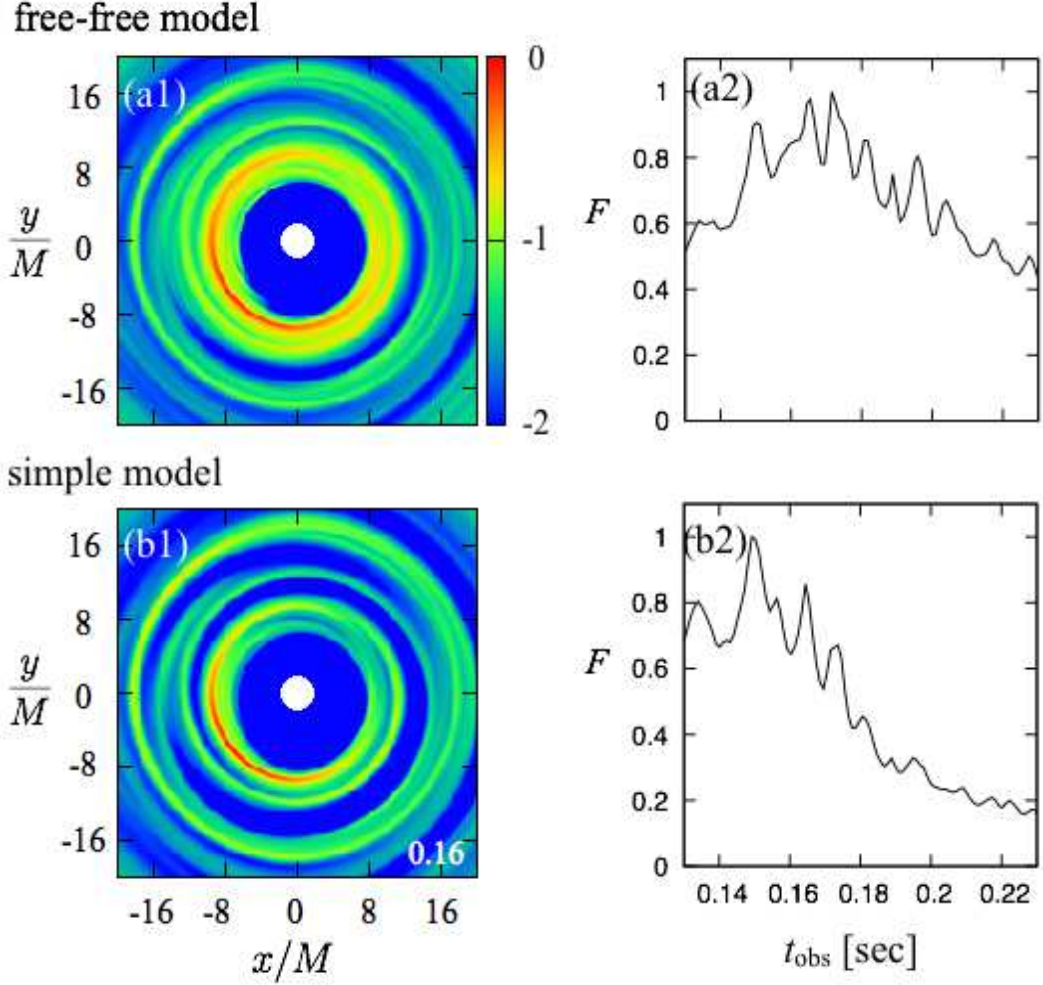} \end{center}
\caption{
Comparison of the dynamical and radiative properties of the gas clouds of the free-free model (upper panels) 
with those of the simple one (lower panels) in the case of the non-rotating black hole.
The left panels show the spatial distribution of the emissivity, where $t=0.16$ sec. 
The definition of the color of panel a1 (b1) is same as that of figure \ref{brems_appendix} (figure \ref{jnu}).
The right panels plot the normalized light curve of each model, where $i=60^\circ$.
}
\label{brems}
\end{figure}

\begin{figure}\begin{center}\includegraphics[width=17cm]{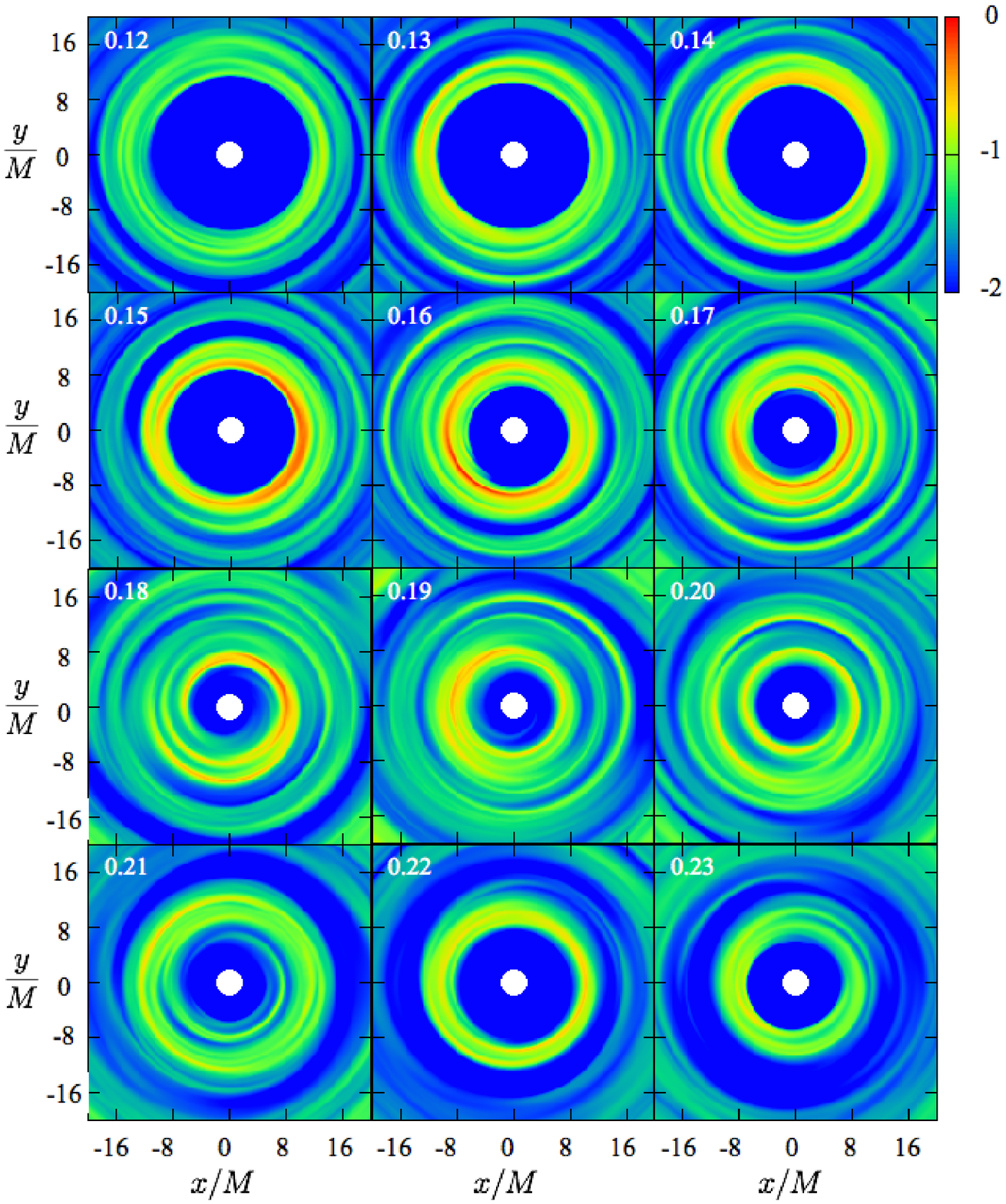} \end{center}
\caption{
The same as figure \ref{jnu}, but for the case of the free-free model.
Here, $\overline{j}(x,y,t)$ reaches its absolute maximum at $(x,y,z,t)=(-5.2M,-7.7M,0.16{\ \rm sec})$
}
\label{brems_appendix}
\end{figure}

In the case of the simple model, we assume the constant electron temperature, $T_e = 10^7$K.
Next, we investigate how the dynamical and radiative properties of arc-shaped gas clouds 
depend on the temperature distribution by using the free-free model (subsection \ref{free}). 
We show in figure \ref{brems} the spatial distribution of the emissivity and light curve of 
the free-free model (upper panels) and those of the simple one (lower pannels), 
where we fix $(a/M, i)=(0,60^\circ)$.
We show in panel a1 the emissivity distribution of the free-free model at the time $t=0.16$ sec, 
and confirm that the shape of the arc-shaped gas cloud is similar to that of the simple one 
(this property is satisfied in the whole time, see, figure \ref{brems_appendix}).
In addition, we compare panel a1 with b1 and find that the low emissivity atmosphere around the arc-shaped gas cloud is thicker than 
that of the simple model, since the hot gas atmosphere exists around the gas cloud.
Therefore, in the case of the free-free model, the whole variation of the emissivity of the gas clouds is gentler than that of the simple model.
In order to show the contribution of the feature to the radiation property, we compare the light curve of the free-free model and that of the simple one 
(see panels a2 and b2).
At the time $0.18<t<0.23$ sec, the overall flux variation of the free-free model is more slowly than that of the simple one, 
and the amplitudes of the short-term light variation are larger.
As a result, we confirm that the dynamical and radiative properties of the gas clouds are not significantly altered from those of the simple model.

\subsection{Results of scattering model: contribution of inverse Compton scattering}
\begin{figure}\begin{center}\includegraphics[width=17cm]{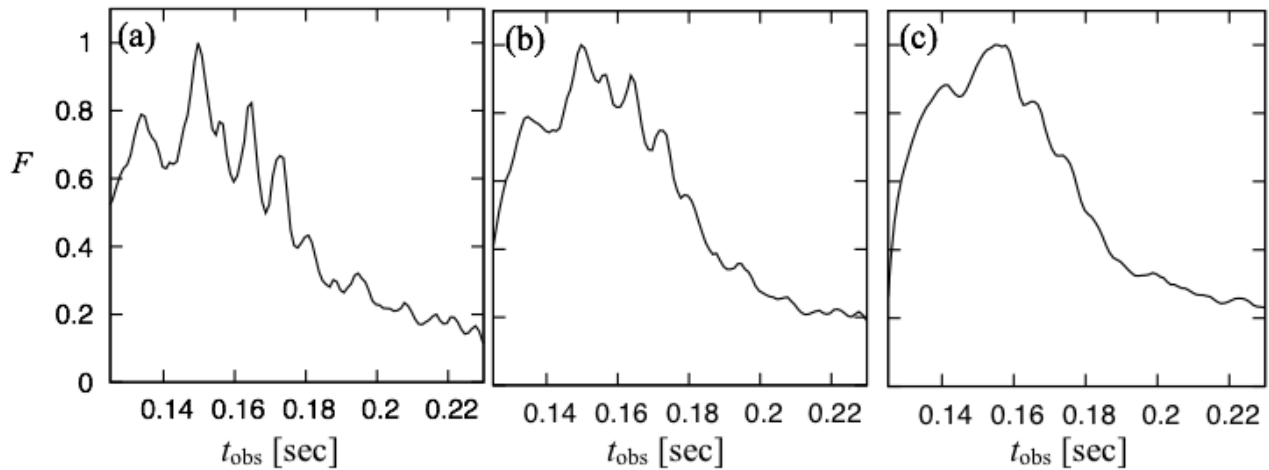} \end{center}
\caption{
Normalized flux variation including the inverse Compton effect for the gas density of the envelope of  
$\rho_{\rm env}=0\ {\rm g\ cm^{-3}}$ (left), $10^{-7}\ {\rm g\ cm^{-3}}$ (middle), and $4.3\times 10^{-7} \ {\rm g\ cm^{-3}}$ (right), respectively.
}
\label{scat_light}
\end{figure}

We investigate the behavior of the short-timescale variation of the flux in the case of the scattering model.
We show in figure \ref{scat_light} the flux variation for 
the no-scattering model [$(\rho,y_{\rm av})=(0,0)$, see panel a], unsaturated-Compton scattering [$(\rho,y_{\rm av})=(1\times 10^{-7},0.3)$, see panel b], 
and critical-saturated Compton scattering [$(\rho,y_{\rm av})=(4.3\times 10^{-7},6.0)$, see panel c].
Here, $y_{\rm crit}$ is the critical y parameter for saturated Comptonization (see problem 7.1 in \citealp{Rybicki1}):
\begin{equation}
y_{\rm crit} = \ln \left(\frac{4T_{\rm env}}{T_{\rm cloud}} \right)=6.0.
\end{equation}
Actually the equation (\ref{scatter_eq}) is invalid in such critical case, but we can examine the qualitative feature of the light variation.
We first confirm that the light curve of the no-scattering model has the same feature as that of the simple one (see panel a of figure \ref{light_u4u}), 
and so we demonstrate the validity of the scattering calculation method. 
Next, we plot the light curve in the case of the unsaturated Compton scattering (panel b), and find that the relativistic light variation can be seen.
The flux variation due to the beaming effect is gentler than the no-scattering model, 
since the electron scattering process spreads the peak width of the relativistic flux variation. 
Even in the case of the critical saturated Compton scattering ($y_{\rm av}= 6.0$), the peak structures are marginally seen (see figure \ref{scat_light}c).
These results show that the fluctuation of the relativistic radiation may be observed even in the case that the inverse Compton scattering affects.

\section{Discussion and future issue}
\subsection{Reproduction of ring-model features}
In this paper, we have confirmed that the accretion flow has the following features near the black hole by using the 3D-GRRMHD simulation data:
\begin{enumerate}
\item 
Arc-shaped gas clouds are intermittently formed near the inner edge of the accretion disk ($R/M=14-16$)
(figures \ref{jnu} and \ref{jnu_high}). 
\item The clouds have nearly the Keplerian orbital velocity and slowly fall to the black hole (figures \ref{light_u4u}b--\ref{light_u4u}d, and \ref{light_u4u_high}b--\ref{light_u4u_high}d). 
\item 
The rotational velocities of the clouds do not depend on the spin value so much. 
This can be understood as follows; in the denominator of the rotational velocity [$v^{\phi} \simeq v_{\rm K} =M^{1/2}/(r^{3/2}+aM^{1/2})$], 
the factor depending on the spin, $aM^{1/2}$, is much smaller than other factor, $r^{3/2}$, at $r/M\approx 10$.

\item 
The light curve has many peaks with short time intervals ($0.01$ sec) due to the beaming effect of 
the non-axisymmetric arc-shaped gas clouds and long time duration ($0.08-0.10$ sec) due to the density variation.
\end{enumerate}
These features are consistent with the key assumptions of our ring model (Paper I). 
Therefore, these results provide support to the situation postulated in Paper I. 
We should note, however, that inverse Compton cooling is not considered in the simulation.
Because of the lack of the cooling process, the gas temperature, $T_{\rm gas}$, is higher than $10^{10}$ K at $R/M<8$, 
and so the arc-shaped gas cloud may be dispersed at the region due to the pressure of the high temperature atmosphere.
In order to understand the accurate behavior of the arc-shaped gas cloud in $R/M<8$, 
we need to incorporate inverse Compton effect to the GRRMHD simulation.

In order to justify the black hole spin measurement proposed in paper I, 
we need to separate the relativistic flux variation from the fluctuation due to the density variation.
Therefore, we must examine whether the light variation due to relativistic effects is distinguished from that due to the density variation,
by performing the simulation for a much longer time during which plenty of gas clouds are successively formed.

\subsection{Observational implications: shot analysis}
In paper I, we argue that the radiation from the gas ring may correspond to the X-ray shots that are flare-like light variations
with sharp peaks.
X-ray shots are flare-like light variations and observed during the low/hard state, whose spectra is characterized by power-law profile (see, e.g., \citealp{Done1}). 
\cite{Negoro1} investigated the variabilities by using the technique of superposed shots, adding plenty of shot profiles 
by aligning their peaks (called “superposed shot” analysis). 
It is known that the superposed shot profile of Cyg X-1 can well be fitted with the sum of two exponential functions: 
one with the time constant of $\sim 0.01$ sec and another of $\sim 0.1$ sec (see paper I and references therein). 
By superposing light curves of the arc-shaped gas-clouds (with different initial phases), 
we expect that the superposed light curve is equivalent to those of a gas ring (see paper I, subsection 5.4).
In the next study, we must calculate the time development of the superposed light curve by performing the long-term simulation, 
and examine the relationship between the short time scale variations ($0.01$ sec) due to relativistic effects and the observational smaller time constant ($\sim 0.01$ sec).

Here, we note that the mass accretion rate of the arc-shaped gas clouds, $\langle \dot{M}_{\rm arc}\rangle$, is consistent with the intensity for the X-ray shot ($\sim 10^{37}\ {\rm erg\ sec^{-1}}$).
We calculate $\langle\dot{M}_{\rm arc}\rangle :=-\int 2\pi Rz \rho u^{r}jd^{3}x\bigg/\int jd^{3}x$, and obtain $\langle \dot{M}_{\rm arc}\rangle = 10^{38-39}\ {\rm erg\ sec^{-1}}$ for the both spin cases.
If we assume that the energy conversion efficiency is $\eta = 10^{-2}-10^{-1}$, the luminosity is $10^{36-38}\ {\rm erg\ sec^{-1}}$ which is on the same order of that of the X-ray shot.

\subsection{Observational implications: HF QPOs}
So far, HF QPOs have been detected in seven sources, such as GRO J1665-40, XTE J1550-564, GRS 1915+105, 
H1743-322, 4U 1630-47, XTE J1859+226, and XTE J1650-500 (\citealp{McClintock1}; \citealp{Remillard4a}).
Three sources (XTE J1859+226, XTE J1650-500, and 4U 1630-47) have single frequency (\citealp{Cui1}; \citealp{Homan1}; \citealp{Remillard4a}). 
The other four sources (GRO J1655-40, XTE J1550-564, H1743-322, and GRS 1915+105) display pairs of HF QPOs with frequencies in a 3:2 ratio (\citealp{Strohmayer1}; \citealp{Remillard1}; \citealp{Homan2}).
Most often, these pairs of QPOs are not always detected.

The Galactic black hole binary system XTE J1550-564 is a typical source of the HF QPOs
(with frequencies of $92, 184$ and $276$ Hz, see \citealp{Remillard1}), 
where  $M=8.4-11.2 M_{\odot}$, $D=5$ kpc, and the binary inclination angle is $70^{\circ}$ (\citealp{Orosz2}).
The outburst found in 1998 was used to examine the disk structure in the very high state when the HF QPOs were observed.
Kubota \& Done (2004) analyzed the X-ray spectra, and reported that following features:
\begin{enumerate}
\item The inner part of the optically thick disk does not reach the radius of the marginally stable orbit ($R/M\approx 10$).
\item The inner disk temperature, $T_{\rm in}$, is relatively cold ($T_{\rm in}\sim 10^{7}$ K).
\item The inner part of the disk is sandwiched by the hot gas regions.
\end{enumerate}
Such features agree well with the simulation data, whereby simulation results indicate that the disk is truncated at $R/M=20-30$, the disk temperature is $\gtrsim 10^{7}$ K, and the disk is sandwiched by the overheated regions (\citealp{Takahashi1}).

From the consuderation, we expect that the relativistic flux-variation of the arc-shaped gas clouds is 
the origin of the HF QPOs.
In the case of a non-rotating black hole, the averaged peak-interval of the arc-shaped gas cloud, $C_1\ (C_2)$, is $\delta t=1.0\times 10^{-2}\ (6.6\times 10^{-3})$ sec [where $1/\delta t=100\ (159)$ Hz], 
and the corresponding radius is $R/M=10\ (7.4)$, where we assume that $\delta t$ is equal to the Keplerian orbital period.
In the case of a high-spin black hole, the averaged peak-interval of the gas cloud is $\delta t=9.7\times 10^{-3}$ sec ($1/\delta t=103$ Hz), and the corresponding radius is $R/M=9.6$.
These frequencies are on the same order of that of the HF QPOs ($100-450$ Hz).

In this study, we do not discuss the power spectra and the commensurability of the frequency of the HF QPOs,
since long-term simulation data, with which statistical analysis can be employed, are unavailable at present.
In order to examine the origin of the pairs of HF QPOs, we need to analyze 
plenty of the formation of arc-shaped gas clouds.

\subsection{Remaining issues}

Finally, we summarize future issues.
\begin{enumerate}
\item In this paper, we assumed that the electron temperature, $T_{\rm e}$, is kept constant everywhere, 
or is equal to the ion temperature (one-temperature plasma). 
In order to accurately calculate the spectral variation and dynamics of arc-shaped gas clouds in the plunging region, 
it is necessary to evaluate the emissivity variation by carefully estimating the temporal and spatial fluctuation of $T_{\rm e}$ 
due to inverse Compton cooling in the original GRRMHD simulation 
(\citealp{Sadowski2a}; \citealp{Sadowski2b}; \citealp{Sadowski1}).
\item In this simulation, the formation of the arc-shaped gas cloud mainly occurred twice (once) for the case of $a/M=0\ (0.9375)$.
In order to examine the major properties of dynamics and radiation of the arc-shaped gas clouds, it is necessary to perform the simulation for a much longer time during which plenty of gas clouds are successively formed.
By using the averaged features of the gas clouds, we need to modify our ring model to the more realistic one.
With such longer simulation data, we accurately evaluate the profile of the power spectra, and may be able to detect the important features of HF QPOs, 
such as the pairs of QPOs with frequencies in a 3:2 ratio.
\item We considered  two spin case ($a/M=0$ and $0.9375$), and reproduced the features of ring model in Paper I.
In order to examine the spin dependence of the light variation, we need to calculate various spin cases.
Then, we must examine the spin dependence of the dynamical features of gas clouds, and test the method of spin measurement proposed in Paper I. 
\end{enumerate}

Numerical computations were carried out on Cray XC30 at the Center for Computational Astrophysics of National Astronomical Observatory of Japan, 
on FX10 at Information Technology Center of the University of Tokyo, and on K computer at AICS. 
This work is supported in part by JSPS Grant-in-Aid for  JSPS Research Fellow (JP17J08829), Scientific Research (C) (17K05383 S. M.,), and Young Scientists (17K14260 H.R.T.).
This research was also supported by MEXT as" Priority Issue on Post-K computer "(Elucidation of the Fundamental Laws and Evolution of the Universe) and JICFuS.






\appendix

In this paper, we transform Kerr-Schild coordinates, ($t_{\rm KS}, r_{\rm KS}, \theta_{\rm KS}, \phi_{\rm KS}$), 
and the four-velocity of the gas element, $u^{\mu}_{\rm KS}$, into Boyer-Lindquist ones, ($t, r, \theta, \phi$), and $u^{\mu}$:
\begin{eqnarray}
t &=& t_{\rm KS}-\int^{r}_{r_{\rm i}}\frac{2Mr}{\Delta}dr,\\ 
r &=& r_{\rm KS},\ \ \ \theta=\theta_{\rm KS}, \\
\phi &=& \phi_{\rm KS}-\int^{r}_{r_{\rm i}}\frac{a}{\Delta}dr, \\
w^{t}&=& w^{t}_{KS}-\frac{2Mr}{\Delta}w^{r}_{KS},\\
w^{r}&=& w^{r}_{KS}, \ \ \ w^{\theta}= w^{\theta}_{KS}, \\
w^{\phi}&=& w^{\phi}_{KS}-\frac{a}{\Delta}w^{r}_{KS}.
\end{eqnarray}
Here, $r_{\rm i}$ is numerical constant, and we choose $r_{\rm i}/M=20$ in order to set $(t,\phi)=(t_{\rm KS},\phi_{\rm KS})$ at the radius $r/M=20$, 
which is the inner edge of the initial torus.

\end{document}